\documentclass[12pt]{article}
\usepackage{fancyhdr}
\usepackage{url}

\usepackage[utf8]{inputenc}

\usepackage{graphicx}

\usepackage{authblk}

\usepackage{booktabs}

\usepackage{xcolor}

\let\oldFootnote\footnote
\newcommand\nextToken\relax

\renewcommand\footnote[1]{%
    \oldFootnote{#1}\futurelet\nextToken\isFootnote}

\newcommand\isFootnote{%
    \ifx\footnote\nextToken\textsuperscript{,}\fi}

 {\fancyhead{}
\fancyfoot[c]{\small{\rule{17.5cm}{1pt}}}}

\begin{document}
\title{\vspace{-2.5cm}
\begin{center}
\textbf{\small{WORKSHOPS REPORT}}\\\vspace{-0.5cm} \rule{17.5cm}{1pt}
\end{center}
\vspace{1cm}\textbf{ECIR 2020 Workshops:\\Assessing the Impact of Going Online}}

\author{
Sérgio~Nunes (INESC TEC and University of Porto),
Suzanne~Little (Insight Centre for Data Analytics, Dublin City University),
Sumit~Bhatia,
Ludovico~Boratto,
Guillaume~Cabanac,
Ricardo~Campos,
Francisco~M.~Couto,
Stefano~Faralli,
Ingo~Frommholz,
Adam~Jatowt,
Alípio~Jorge,
Mirko~Marras,
Philipp~Mayr,
Giovanni~Stilo

}

\date{
{\color{blue}
Preprint of the report submited to ACM SIGIR Forum June edition\break
See \url{http://sigir.org/forum/}\break
\newline
\today}}

\maketitle
\thispagestyle{fancy}

\section{Introduction}
ECIR 2020\footnote{\url{https://ecir2020.org/}} was one of the many conferences affected by the COVID-19 pandemic. The Conference Chairs decided to keep the initially planned dates (April 14-17, 2020) and move to a fully online event. In this report, we describe the experience of organizing the ECIR 2020 Workshops in this scenario from two perspectives: the workshop organizers and the workshop participants. We provide a report on the organizational aspect of these events and the consequences for participants. Covering the scientific dimension of each workshop is outside the scope of this article.

The results presented here are based on two online questionnaires involving the participation of more than 60 people: one sent to all workshop organizers and another one sent to all workshop paper authors and workshop participants. The former is presented in Section~\ref{section:organizers} to describe the workshop organizers' feedback. The latter is introduced in Section~\ref{section:participants} to describe the participants' perspective. Our general objective is to get a sense of the feeling of both stakeholders in this first-time experience. Questions such as \emph{``What was gained/lost?''}; \emph{``What was harder/easier?''}, \emph{``What worked?''}; \emph{``What didn't work?''} among many others will be dissected. We conclude this article by highlighting the main recommendations and lessons learned from this online experience. We believe this to be of potential importance for researchers interested in organizing similar events in the future.
\break
\section{Workshop Organization}~\label{section:organizers}
\vspace{-1cm}
\subsection{Overview}
ECIR 2020 Workshops were held on April 14, 2020 (Tuesday) adopting Lisbon time (CEST+1), from 09h00 to 18h00. The first call for workshops was issued on July 22, 2019. We received a total of six proposals, one was later withdrawn, and four were accepted -- three as full day events, one half-day. Workshop notifications were sent on October 1, one month after the submission deadline. On March 11, the decision to move to a fully online event was announced. We present a summary of the accepted ECIR 2020 Workshops in Table~\ref{table:workshops}.

\begin{table}[]
\centering
\caption{ECIR 2020 Workshops.}
\label{table:workshops}
\begin{tabular}{lccc}
\hline
\textbf{Workshop}     & \textbf{Submissions} & \textbf{Accepted}           & \textbf{Participants} \\ \hline
BIAS (full-day)  & 44 & 11 long; 6 short               & \textgreater~70 \\
BIR (full-day)   & 9  & 8 full              & \textgreater~70        \\
SIIRH (half-day) & 7  & 3 full; 3 short & \textgreater~50 \\
Text2Story (full-day) & 20                        & 10 full; 2 position; 1 demo & \textgreater~70       \\ \hline
\end{tabular}
\end{table}

The ECIR 2020 Organization adopted Zoom as the official platform for synchronous video conferencing. A dedicated Zoom room with password access was configured for each workshop and the details provided to the workshop organisers. A few days before the event, a draft version of ACM’s best practices guide for virtual conferences~\cite{ACMVirtualConferences} was published and shared with the participants. The ECIR 2020 Organisation also created a slack workspace for use during the event and a dedicated slack channel for workshop organizers was created. This was used in particular to connect with technical support during the workshops. The specific details of how each workshop was organized are described next.

\paragraph{BIAS 2020~\cite{BIAS2020}} was the 1st edition of the International Workshop on Algorithmic Bias in Search and Recommendation\footnote{\url{http://bias.disim.univaq.it/}}. BIAS was organized as a full-day event, chaired by Ludovico~Boratto (EURECAT, Centre Tecnològic de Catalunya), Mirko~Marras (University of Cagliari), Stefano~Faralli (University of Rome Unitelma Sapienza), and Giovanni~Stilo (University of L’Aquila). The workshop aimed at collecting novel ideas to detect, measure, characterize, and mitigate bias in data and algorithms underlying search and recommendation applications. The goal was to provide a common ground for researchers and practitioners working in this area. The workshop had more than 70 participants. The keynote speaker -- Prof. Chirag Shah from the University of Washington (USA) -- highlighted how bias, especially in relation to search and recommender systems, causes material problems for users, businesses, and society at large. The scientific program included demo and paper presentations and a final discussion. The papers covered topics from search and recommendation in online dating, education, and social media, over the impact of gender bias in word embeddings, to tools that allow exploration of bias and fairness on the Web. The event concluded with a discussion session to highlight open issues, research challenges, and briefly summarize the outcomes of the workshop. The workshop proceedings volume is going to be published soon in Springer's ``Communications in Computer and Information Science'' series.

\paragraph{BIR 2020~\cite{BIR2020}} was the 10th edition of the International Workshop on Bibliometric-enhanced Information Retrieval\footnote{\url{https://sites.google.com/view/bir-ws/bir-2020}}. BIR was organized as a full-day event and chaired by Guillaume~Cabanac (Université de Toulouse), Ingo~Frommholz (University of Bedfordshire), and Philipp~Mayr (GESIS -- Leibniz-Institute for the Social Sciences). BIR tackles issues related to academic search, at the crossroads between Information Retrieval, Natural Language Processing and Bibliometrics. The keynote speaker -- George Tsatsaronis, VP Data Science at Elsevier -- addressed the topic of Metrics and Trends in Assessing the Scientific Impact. Eight papers were given live (no pre-recorded videos), each one followed by a Q\&A session and greetings for the 10th anniversary edition from senior scholars in our field. The proceedings are published in open access at CEUR-WS~\cite{BIR2020proc}.

\paragraph{SIIRH 2020~\cite{SIIRH2020}} was the 1st edition of the International Workshop on Semantic Indexing and Information Retrieval for Health from Heterogeneous Content Types and Languages. SIIRH was organized as a half-day event and chaired by Francisco~Couto (LASIGE, Universidade de Lisboa) and Martin~Krallinger (BSC-CNS). The workshop included an open session covering novel contributions on IR technologies for heterogeneous health-related content open to multiple languages with a particular interest in the exploitation of structured controlled vocabularies and entity linking for document indexing and semantic search applications. The program committee was composed of 46 international researchers and selected three full papers and three short papers. A MESINESP/Plan TL Session was also part of the workshop for presentation and discussion of current work and new advances in the MESINESP shared task or other medical IR, QA or text categorization evaluation campaigns, as well as the exploitation of evaluation settings and data collections generated through these kind of community evaluation efforts. The Open Session included two keynote talks, one by Dr. Cher Han Lau, entitled CoronaTracker: A framework for managing and tracking data during crisis, and another by Lucy Lu Wang and Kyle Lo, entitled The COVID-19 Open Research Dataset. The MESINESP/Plan TL Session included one keynote talk, by George Paliouras, entitled BioASQ: The challenge and the community of biomedical semantic indexing and question answering. Recordings of all talks are publically available in YouTube\footnote{\url{https://www.youtube.com/playlist?list=PL6RYRv3A1tLwpD4aTbSVraUZNITbkRxZd}} and the preliminary proceedings are also publically available online\footnote{\url{https://drive.google.com/open?id=1-sF_0R3uGinq5ybcAM5H54k5yujJE8o6}}.

\paragraph{Text2Story 2020~\cite{Text2Story2020}} was the 3rd edition of the International Workshop on Narrative Extraction from Texts\footnote{\url{http://text2story20.inesctec.pt}}, chaired by Ricardo~Campos (INESC TEC; Ci2 – Smart Cities Research Center, Polytechnic Institute of Tomar), Alípio~Jorge (INESC TEC and FCUP - University of Porto), Adam~Jatowt (Kyoto University) and Sumit~Bhatia (IBM Research AI). It was organized as a full-day event and consisted of presentations of ten regular research papers, two position papers and one demonstration paper. The papers presented at the workshop\footnote{\url{https://text2story20.inesctec.pt/text2story2020_slides.zip}}\footnote{\url{https://www.youtube.com/channel/UC8ysv2TEpQZz_-v5T5nZSSA/featured}}, were made available at CEUR~\cite{Text2Story2020proc} and cover diverse aspects of the narrative extraction problem ranging from applications of deep learning to narrative extraction, sentiment analysis, event coreference resolution to applications in different domains such as legal text, news and social media. The program also included two invited keynote talks. Sebastião Miranda, head of development at Priberam (Portugal), talked about monitoring multi-lingual, multimedia sources of information to discover relevant stories, events, and topics of interest to a specific user. Mark Finlayson, assistant professor at the School of Computing and Information Sciences of the Florida International University (USA), presented an overview of recent advances in Natural Language Processing applied to the task of narrative analysis in text. 

A more detailed overview of the BIAS, BIR, and Text2Story workshops can be found in this edition of SIGIR Forum.

\subsection{Feedback from Workshop Organizers}
To collect feedback from workshop organizers, an online questionnaire was prepared and shared on May 2, 2020 and kept open until the end of May 5. Out of a total of 13 organizers, we received responses from 9.

The responses to the questionnaire revealed that most of the organizers had a very positive experience that generally exceeded their expectations after knowing that the event would be remote. With a view to the future, while the majority of the respondents stated that they would be willing to organize future events similarly, a significant number of them also indicated that they would do so ``Only if needed''. Overall, every respondent expressed their support for remote participation in the future, although some did state that this should be allowed only for specific cases.

Regarding the effort required for the organization of the event, the respondents had very different experiences ranging from ``the same effort as in-person'', ``more than in-person'' to ``less than in-person''. As a preventive measure, most of the organizers requested the authors to provide pre-recorded videos of their presentations in case of some unforeseen last-minute technical glitches. Regarding the usefulness of the platforms made available to the organizers, almost all the respondents found Zoom to be a crucial tool required for organizing the online event, and some of the respondents also found the Zoom Chat feature helpful for interaction with the participants. We also note that in addition to the live and synchronous communication media (Zoom Chat and Slack), traditional email also played a crucial role in planning and communication between the organizers and participants before, during, and after the event.

A more detailed analysis of the results can be found below.

\subsubsection{Analysis of closed and ratings questions}
From the respondents, all except one (8) had previous experience in organizing scientific events. To the question \emph{``What is your overall feeling about how the workshop went?''}, 4 answered ``Very Happy'' (44\%) and 5 answered ``Happy'' (56\%). No respondents reported ``Unhappy'' nor ``Very Unhappy''. To the question \emph{``How does it compare with your previous expectations (after knowing the event would be remote)?''} (8 answers), 2 answered ``Much better than expected'' (25\%) and 6 answered ``Better than expected'' (75\%). No answers of ``Worse than expected'' or ``Much worse than expected'' were reported.

To the question \emph{``Would you organize this event again in the future in a similar online model?''}, 5 respondents answered ``Yes'' and 4 answered ``Only if needed''. None of the respondents answered ``No'' to this. To the question \emph{``For scientific events conducted in person, how do you evaluate the importance of allowing remote participations?''}, 4 respondents answered ``Should be standard practice and allowed for anyone'' and 5 answered ``Should be possible only for specific cases''. None of the respondents selected ``Should be avoided'' or ``Should not be allowed''. This indicates that in future, a hybrid organization model for workshops could be used where, for example, presentations are given by paper authors in person for both the on-site and remote audience at the same time.

\begin{figure}
\centering
\includegraphics[width=0.75\textwidth]{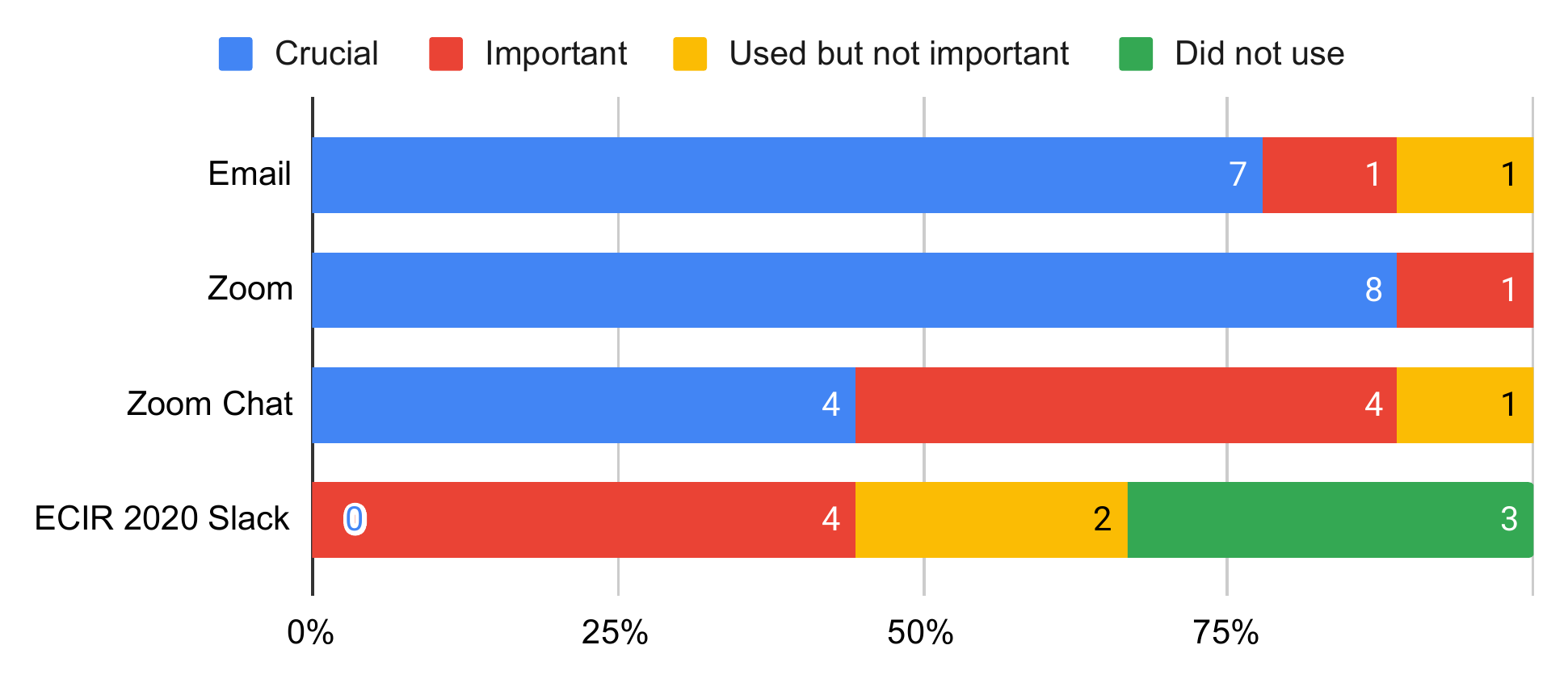}
\label{figure:organizers-tools}
\caption{Usefulness of the tools as reported by organizers.}
\end{figure}

Coming to the questions related to organization, for the question \emph{``How do you evaluate the overall effort associated with the organization of the workshop online?''}, the majority answered ``Different, but the same effort as in person'' (5), with 2 respondents answering ``More than in person'' and 2 answering ``Less than in person''. For the question \emph{``Did you use pre-recorded videos or synchronous presentations?''}, the answers were 5 ``Both, depending on conditions'' and 4 ``Synchronous''. Figure~\ref{figure:organizers-tools} presents the results to the question \emph{``Rate the usefulness of the tools used''}.

\subsubsection{Analysis of comments from open questions}
\begin{quote}
    ``Undoubtedly, having the workshop online allowed us to reach a broader audience in terms of participants. We believe that having conferences online in the future will also enable a higher number of submissions from a number of researchers, which for different reasons, anticipate not to attend a conference.''
\end{quote}

Positive comments from the workshop organisers emphasized the higher levels of participation and the diversity of world-wide attendees due to not having to travel and the subsequent reduced costs. BIR, which had run previously, saw an increase in the number of participants from about 30 to almost 100 while other workshops also received very strong attendance -- almost double the expected numbers in one case. Comments on the use of Zoom suggest that it handled the workload well and removed some of the issues with a projector and shared computers that occur sometimes in a physical setting. Zoom presentations also meant that the viewers could all clearly see the slide and post questions on the chat ahead of the Q\&A session. Subsequent questions could be clearly heard and understood by participants. However not all the organisers felt that this was a positive experience with some respondents unable to identify any specific positive aspects to running an online workshop.

Comments from the workshop organisers on the negative aspects of the online workshops identified a number of specific challenges and issues. A common response highlighted the lack of interactivity and connection between participants -- ``limited feedback'', ``personal interaction was totally missing'', ``chatting between participants'', ``start[ing] a profitable discussion'', ``[no] quick chats with colleagues''. Two respondents also said they just missed human touch and hugs and one commented that although the workshop ran well they ``missed the personal contact a lot''. Attempts to encourage chatting using breakout rooms were not successful, especially where attendees did not know each other previously. The natural result was the lack of atmosphere -- ``things were colder'', ``[not able to] joke around with audience'', ``[no] stimulating interaction'', ``[no] eye contact'', ``[no] spontaneity''. The organisers also commented that as the chairs or session chairs they needed a much higher level of concentration and intensity through the event to manage the platform and bilateral discussions. Apart from a specific instance of Zoombombing~\cite{Zoombombing} in one workshop, there were also some individual network issues that disrupted sessions, challenges around time zones and communicating the registration policy.

\begin{quote}
    ``I felt the online worked as a barrier to have people engaged when compared to a regular conference.''
\end{quote}

\begin{quote}
    ``All in all, the entire process is pretty much demanding and tireless. As an organizer, I didn't find it much [sic] rewarding in the sense that I didn’t have time to look at the presentations.''
\end{quote}

\subsubsection{Recommendations from the organisers}
Organisers were asked to give their recommendations for best practise in running an online workshop. Key considerations in preparation include keeping the programme short and dynamic as the required level of concentration is much greater; communicating detailed instructions via email both prior to and just before the event starts and planing a test or practise meeting for presenters are just some of the listed ideas. Pre-recording talks and making them available via a channel like YouTube was also useful. A good suggestion was to involve student volunteers to help with managing communications during the event. We all learned that participants should be explicitly warned against (accidental) sharing of Zoom hyperlinks (which include the room’s password) via social media as this likely led to the Zoombombing incident of one workshop. Furthermore, when using Zoom, organisers should consider activating the ``waiting room'' functionality to better manage the participants.

During the event, the moderator and session chair should proactively mute microphones during talks and then manage the questions either via the chat or through attendees raising hands and being asked to talk. Organisers made some suggestions to try and incentivise feedback and participation including asking people to show their face, suggesting to open the microphones to give a round of applause after talks, introducing the audience (character and number) and using Zoom Chat to contact people.

A fully online workshop can also have different opportunities post event. Organisers suggest that you can follow up by publishing the videos, slides or talks, using a chat option like Slack to have further discussions and trying to use social media and the workshop website to promote a sense of community. This is particularly difficult when participants haven’t met physically. Conducting post-event surveys or other contact with presentors and participants is key.

\section{Workshop Participation}~\label{section:participants}
To collect feedback from workshop participants, an online questionnaire was sent to all workshop paper authors and participants on May 6, 2020 and was kept open until the end of May 8. A total of 47 answers were received and the results are presented below.

\subsection{Analysis of closed and ratings questions}
Approximately 50\% of all respondents were not paper authors (24 answers) and for 85\% (40 answers) this was their first participation in the respective workshop. For the question \emph{``Would you have attended the workshop in Lisbon (if this pandemic had not happened)?''}, 19 answered ``Yes'' (43\%), 20 answered ``No'' (40\%), and the remaining 8 answered ``Not sure''.

From all respondents, a large majority reported previous participation in scientific events (87\%, 41 answers). In the answer to the question \emph{``How do you compare this experience with your previous participations in scientific events?''}, 12 answered ``Better. The remote model improved my experience'' (26\%), 10 answered ``Similar. The remote model had no relevant impact'' (21\%), 18 answered ``Worse. The remote model negatively impacted the event'' (38\%), and 7 selected ``No answer. I have no relevant previous experience to compare with'' (15\%).

\begin{figure}[h]
\centering
\includegraphics[width=0.75\textwidth]{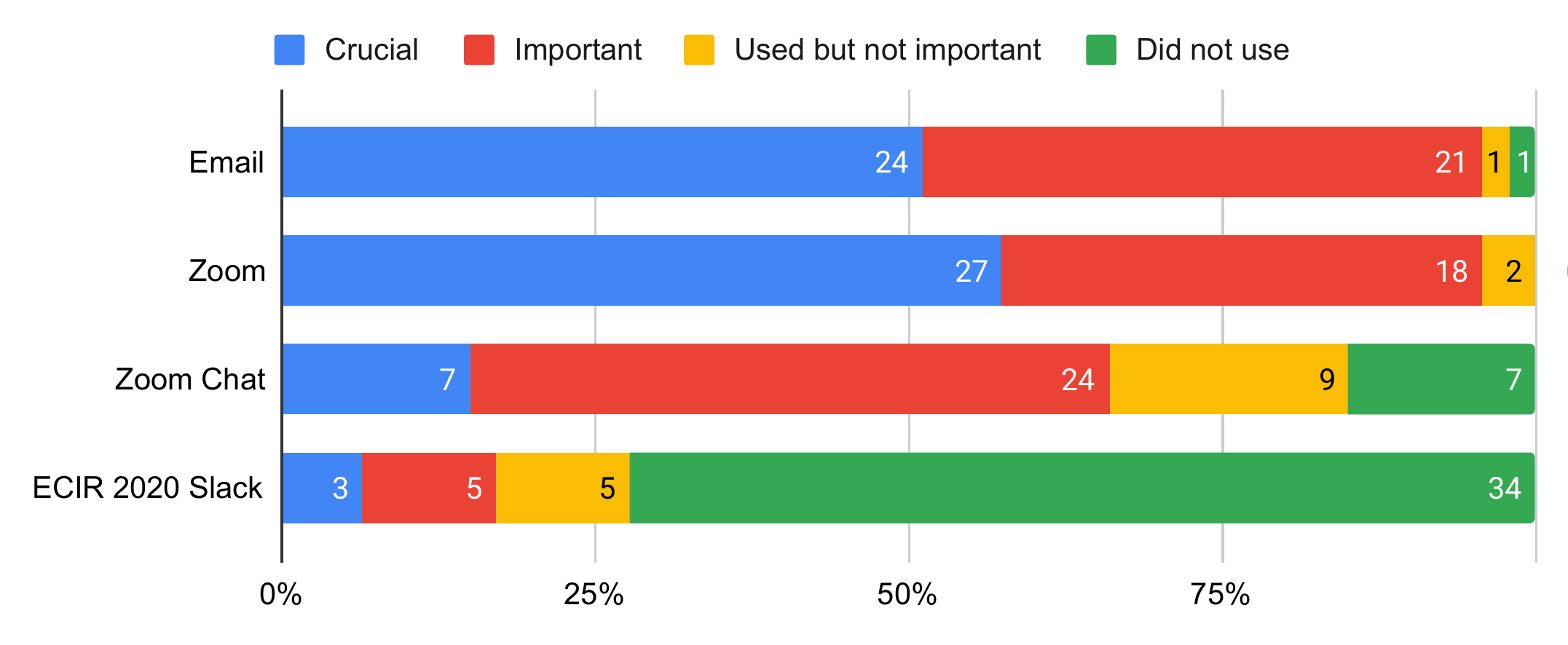}
\label{figure:participants-tools}
\caption{Usefulness of the tools as reported by participants.}
\end{figure}

For the question \emph{``How does it compare with your previous expectations (after knowing the event would be online)?''}, 12 answered ``Much better than expected'' (26\%), 31 answered ``Better than expected'' (66\%), 4 answered ``Worse than expected'' (9\%), and no one answered ``Much worse than expected''. Figure~\ref{figure:participants-tools} presents the results to the question \emph{``Rate the usefulness of the following tools''}. We can observe that while Zoom was obviously helpful along with email communication Zoom Chat and Slack were considered as less crucial, although Zoom Chat was still seen as an important communication channel.

\begin{figure}[h]
\centering
\includegraphics[width=0.75\textwidth]{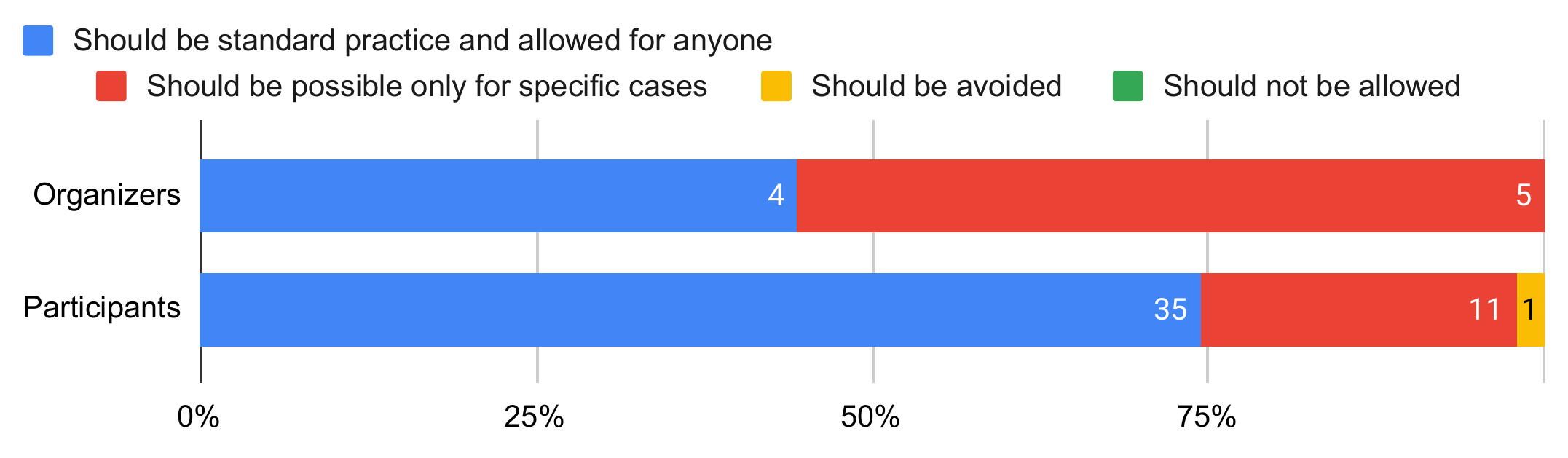}
\label{figure:remote-participation}
\caption{Importance of allowing remote participations as reported by organizers and participants.}
\end{figure}

When asked \emph{``For scientific events conducted in person, how do you evaluate the importance of allowing remote participations?''}, a majority of 35 respondents answered ``Should be standard practice and allowed for anyone'' (75\%), 11 answered ``Should be possible only for specific cases'' (23\%), 1 answered ``Should be avoided'' and no one answered ``Should not be allowed''. This suggests that conferences and workshops could consider opening up in one way or other for remote participation. In Figure~\ref{figure:remote-participation} is a comparison between the feedback from organizers and participants. It is interesting to note the differences depending on the role assumed.

\subsection{Analysis of comments from open questions}
Workshop attendees who were not paper authors (26 respondents) commented positively on the improved possibility and ease of attending, reduced travel costs, the quality of the presentations, the effectiveness of the platform and the ease of asking questions as well as comments on unexpected benefits like the relaxed dress code and the option to screenshot interesting slides to follow up. The majority of negative comments referenced the lack of personal interactions and socialising, missing coffee breaks and networking with colleagues in general. The specific incident where one workshop was invaded was also a negative experience for attendees. Some found it difficult to have physical spaces without disruption due to the isolation in their home or other work commitments, and others faced internet connectivity challenges. 

Responses of attendees who were authors or presenters of papers (16 responses) agreed that it was much easier and cheaper to attend and one noted the particular benefit for students to participate and the opportunity to present to a wider audience. It also meant that more than one author per paper could participate. They also liked the Q\&A options, especially the chat and options to screen share. A number of respondents commented positively on the digital presentation experience and the option to pre-record their presentation that was provided. Amusingly one noted that it was easier to present without seeing the audience while another liked the option to take a break without being noticed. 

Missing personal contact, networking and lack of interaction also came up most frequently in the negative comments from paper authors. In particular the lack of extended scientific exchange and socialisation was a prevailing concern. One commented on the difficulty in giving an engaging presentation when you cannot see the participants and the lack of eye contact. Some felt that there were too many emails, different credentials and schedule changes that made participating challenging. They also found it more difficult to focus for a full day. Many comments also noted technical difficulties including sound issues, internet access and a good suggestion was made to invest in a cheap microphone if you have to present in an online event. A comment specific to workshops was that it felt more difficult to engage with the audience when you were presenting cross-discipline work.

\begin{quote}
    ``I think I really missed networking with the researchers and experts from the workshop''
\end{quote}

\section{Conclusion}
In conclusion, the overall experience of running the ECIR 2020 workshops as synchronous, fully online events was generally found to be a positive one. Approximately 25\% of both organizers and participants who responded, reported it was ``Much better than expected''. The major advantage was felt to be the increase in both the number of attendees and the breadth, enabling those who may ordinarily have not been able to travel to participate. Especially for workshops with small budgets, online events facilitate the process of including keynote speakers. Online presentations, coupled with the requirement or option to send pre-recorded presentations have potential to alleviate the issue of no-shows which tend to be the concern of many organizers in the past. There were individual technical challenges (sound, connectivity, etc.) and one security incident but the major disadvantage was the lack of social interaction, of scientific networking and the ability to connect with people and discuss topics. Finding solutions to mitigate this decrease in interaction would greatly improve the online workshop experience.

ECIR 2020 was a fully online event, with more than 1000 registrees. In this report, we presented the first-hand experience of both ECIR 2020 Workshop organizers and participants, collected using two online questionnaires. While this was setup to deal with a health emergency situation, we think that the options for fully, or hybrid, online events will increase in the future. Not only due to some of the aforementioned advantages but also as proactive actions to reduce academia carbon footprint associated with air travel~\cite{CarbonFootprint}.

\bibliographystyle{acm}
\bibliography{references}

\begin{thebibliography}{1}

\bibitem{ACMVirtualConferences}
{\sc {Association for Computing Machinery}}.
\newblock {Virtual Conferences - A Guide to Best Practices (Report version 1.1
  from April 13, 2020). ACM Presidential Task Force.}
\newblock \url{https://www.acm.org/virtual-conferences}, 2020.
\newblock [Online; accessed 11-May-2020].

\bibitem{BIAS2020}
{\sc Boratto, L., Marras, M., Faralli, S., and Stilo, G.}
\newblock {International Workshop on Algorithmic Bias in Search and
  Recommendation (Bias 2020)}.
\newblock In {\em Advances in Information Retrieval - 42nd European Conference
  on {IR} Research, {ECIR} 2020, Lisbon, Portugal, April 14-17, 2020,
  Proceedings, Part {II}\/} (2020), J.~M. Jose, E.~Yilmaz, J.~Magalh{\~{a}}es,
  P.~Castells, N.~Ferro, M.~J. Silva, and F.~Martins, Eds., vol.~12036 of {\em
  Lecture Notes in Computer Science}, Springer, pp.~637--640.

\bibitem{BIR2020}
{\sc Cabanac, G., Frommholz, I., and Mayr, P.}
\newblock {Bibliometric-Enhanced Information Retrieval 10th Anniversary
  Workshop Edition}.
\newblock In {\em Advances in Information Retrieval - 42nd European Conference
  on {IR} Research, {ECIR} 2020, Lisbon, Portugal, April 14-17, 2020,
  Proceedings, Part {II}\/} (2020), J.~M. Jose, E.~Yilmaz, J.~Magalh{\~{a}}es,
  P.~Castells, N.~Ferro, M.~J. Silva, and F.~Martins, Eds., vol.~12036 of {\em
  Lecture Notes in Computer Science}, Springer, pp.~641--647.

\bibitem{BIR2020proc}
{\sc Cabanac, G., Frommholz, I., and Mayr, P.}, Eds.
\newblock {\em {Proceedings of the 10th International Workshop on
  Bibliometric-enhanced Information Retrieval co-located with 42nd European
  Conference on Information Retrieval, BIR@ECIR 2020, Lisbon, Portugal, April
  14th, 2020 [online only]}\/} (2020), vol.~2591 of {\em {CEUR} Workshop
  Proceedings}, CEUR-WS.org.

\bibitem{Text2Story2020}
{\sc Campos, R., Jorge, A., Jatowt, A., and Bhatia, S.}
\newblock {The 3\({}^{\mbox{rd}}\) International Workshop on Narrative
  Extraction from Texts: Text2Story 2020}.
\newblock In {\em Advances in Information Retrieval - 42nd European Conference
  on {IR} Research, {ECIR} 2020, Lisbon, Portugal, April 14-17, 2020,
  Proceedings, Part {II}\/} (2020), J.~M. Jose, E.~Yilmaz, J.~Magalh{\~{a}}es,
  P.~Castells, N.~Ferro, M.~J. Silva, and F.~Martins, Eds., vol.~12036 of {\em
  Lecture Notes in Computer Science}, Springer, pp.~648--653.

\bibitem{Text2Story2020proc}
{\sc Campos, R., Jorge, A.~M., Jatowt, A., and Bhatia, S.}, Eds.
\newblock {\em {Proceedings of Text2Story - Third Workshop on Narrative
  Extraction From Texts co-located with 42nd European Conference on Information
  Retrieval, Text2Story@ECIR 2020, Lisbon, Portugal, April 14th, 2020 [online
  only]}\/} (2020), vol.~2593 of {\em {CEUR} Workshop Proceedings},
  CEUR-WS.org.

\bibitem{SIIRH2020}
{\sc Couto, F.~M., and Krallinger, M.}
\newblock {Proposal of the First International Workshop on Semantic Indexing
  and Information Retrieval for Health from Heterogeneous Content Types and
  Languages {(SIIRH)}}.
\newblock In {\em Advances in Information Retrieval - 42nd European Conference
  on {IR} Research, {ECIR} 2020, Lisbon, Portugal, April 14-17, 2020,
  Proceedings, Part {II}\/} (2020), J.~M. Jose, E.~Yilmaz, J.~Magalh{\~{a}}es,
  P.~Castells, N.~Ferro, M.~J. Silva, and F.~Martins, Eds., vol.~12036 of {\em
  Lecture Notes in Computer Science}, Springer, pp.~654--659.

\bibitem{CarbonFootprint}
{\sc Quinton, J.~N.}
\newblock Cutting the carbon cost of academic travel.
\newblock {\em Nature Reviews Earth {\&} Environment 1}, 1 (Jan 2020), 13--13.

\bibitem{Zoombombing}
{\sc {Wikipedia contributors}}.
\newblock {Zoombombing} -- {W}ikipedia{,} the free encyclopedia.
\newblock
  \url{http://en.wikipedia.org/w/index.php?title=Zoombombing&oldid=955836412},
  2020.
\newblock [Online; accessed 11-May-2020].

\end{thebibliography}

\end{document}